\documentclass{kluwer}    % Specifies the document style.

\usepackage{epsf}

\def\degree{\relax \ifmmode ^o \else $^o$\fi}

\def\muo{\relax \ifmmode \mu_{\rm 0}\else $\mu_{\rm 0}$\fi}
\def\mue{\relax \ifmmode \mu_{\rm e}\else $\mu_{\rm e}$\fi}
\def\muave{\relax \ifmmode \langle\mu\rangle_{\rm e}\else $\langle\mu\rangle_{\rm e}$\fi}
\def\muaved{\relax \ifmmode \langle\mu\rangle_{\rm e}^{\rm D}\else $\langle\mu\rangle_{\rm e}^{\rm D}$\fi}
\def\magtot{\relax \ifmmode M_I\else $M_I$\fi}
\def\magd{\relax \ifmmode M_{I}^{\rm D}\else $M_{I}^{\rm D}$\fi}
\def\re{\relax \ifmmode r_{\rm e}\else $r_{\rm e}$\fi}
\def\red{\relax \ifmmode r_{\rm e}^{\rm D}\else $r_{\rm e}^{\rm D}$\fi}
\def\res{\relax \ifmmode r_{\rm e*}\else $r_{\rm e*}$\fi}
\def\h{\relax \ifmmode h\else $h$\fi}

\def\vmax{\relax \ifmmode V_{\rm max}\else $V_{\rm max}$\fi}

\def\aap{A\&A}
\def\aaps{A\&AS}
\def\apj{ApJ}
\def\apjs{ApJS}
\def\aj{AJ}
\def\mnras{MNRAS}

\begin{document}
\begin{article}
\begin{opening}         
\title{The space density of spiral galaxies as function of their
scale size, surface brightness and luminosity}

\author{Roelof S. \surname{de Jong}%
\thanks{Hubble Fellow}
\email{rdejong@as.arizona.edu}
}
\institute{Steward Observatory, 933 N.\ Cherry Ave., Tucson, AZ 85721, USA}

\author{Cedric \surname{Lacey} \email{Cedric.Lacey@durham.ac.uk}}
\institute{%TAC, Juliane Maries Vej 30, DK-2100 Copenhagen O, Denmark\\
Univ.\ of Durham, Dept.\ of Physics, South Road, Durham DH1 3LE, UK}

\runningauthor{Roelof de Jong \& Cedric Lacey}
\runningtitle{The space density of spiral galaxies}
%\date{April 15, 1993}

 \begin{abstract}
 The local space density of galaxies as a function of their basic
structural parameters --luminosity, surface brightness and scale size--
is still poorly known.  Our poor knowledge is the result of strong
selection biases against low surface brightness {\em and} small scale
size galaxies in any optically selected sample.  We derive bivariate
space density distributions by correcting a sample of $\sim$1000 local
Sb-Sdm spiral galaxies for its selection effects.  We present a
parameterization of these bivariate distributions, based on a Schechter
type luminosity function and a log-normal scale size distribution at a
given luminosity.  We next calculate the bivariate distributions as
function of redshift using the Hubble Deep Field, and conclude that at
higher redshift there is a decrease in space density of luminous, large
scale size galaxies, but the density of smaller galaxies stays nearly
the same. 
 \end{abstract} 
 \keywords{spiral galaxies, selection effects, structural parameters}

\end{opening}           

\section{Introduction}

In the last few decades many papers have been devoted to the
determination of the luminosity function (LF), the central surface
brightness distribution and, to some lesser extent, the scale size
distribution of galaxies.  These distribution determinations can
actually not be separated due to the limits on the survey material on
which these investigations are based.  Any galaxy LF is only valid to
the surface brightness limit of the survey.  Any surface brightness
distribution has validity limits depending on the scale size and/or
magnitude limits of the survey.  In this paper we investigate the galaxy
bivariate distribution functions using combinations of luminosity,
surface brightness and scale size. 

Bivariate distribution functions have two important applications.  First
of all, using bivariate functions is the proper way to compare samples
with different selection functions, especially when comparing samples at
different redshifts.  Secondly, bivariate distribution functions provide
excellent checks for galaxy formation and evolution theories.  Any
complete galaxy formation theory will have to explain the bivariate
distribution functions of structural galaxy parameters. 

%In this paper we will first discuss selection bias and corrections in
%some detail and than apply these correction to a local sample of about
%a 1000 galaxies to calculate bivariate distribution functions. We
%derive a functional description of the bivariate distributions based
%on hierarchical galaxy formation model and fit this function
%to the data. Finally, we look at the evolution of bivariate functions
%using the data from the Hubble Deep Field (HDF).

%----------------------------------------------------------------------
\section{Visibility Correction}
\label{viscor}

Our aim is to determine the average space density of galaxies with
certain properties in an average volume in the universe.  Most
field galaxy samples are not volume limited, but for instance magnitude
or diameter limited.  Not all galaxies have the same luminosity or
diameter and therefore each galaxy will have a different distance range
where it can be placed before dropping out of the sample due to the
selection criteria.  The maximum volume where a galaxy can be seen and
included into the sample (\vmax) goes as the distance limits cubed,
which results in galaxy samples being dominated by intrinsically bright
and/or large galaxies, because they have a hugely larger visibility
volume (Davies et al.~\citeyear{Dav94}; McGaugh et
al.~\citeyear{McG95}).  We have to correct our sample for this selection
effect to compute the true space density from the observed distribution. 

In this paper we use one of the most simple selection effect correction
methods available, the \vmax\ correction method.  Each galaxy gets a
weight inverse proportional to its maximum visibility volume set by the
selection limits.  For a sample with upper ($D_{\rm max}$) and lower
($D_{\rm min}$) major axis diameter limits this leads to
 \begin{equation}
V_{\rm max} = \frac{4\pi}{3}d^3 ((\frac{D_{\rm maj}}{D_{\rm max}})^3
 -(\frac{D_{\rm maj}}{D_{\rm min}})^3)
\label{vmax}
 \end{equation}
 with $d$ the distance and $D_{\rm maj}$ the major axis diameter of the
galaxy.  Other limits, like redshift, magnitude or sky fraction limits,
that would limit \vmax, can trivially be taken into account as well.
For a sample of galaxies which is complete to within the selection
limits we can now define an estimator of the bivariate distribution
function in parameters \hbox{$x$ and $y$}: 
 $
 \Phi(x,y) \approx \frac{1}{\Delta x\Delta y} \sum_{i}^N 
	\frac{\delta^i}{\vmax^i}
 $,
where $i$ is summed over all $N$ galaxies and $\delta^i=1$ if ($x_i$,$y_i$) of
a galaxy is in the ($x\pm \Delta x/2,y\pm \Delta y/2$) bin range,
otherwise 0.

The parameters used to select the galaxy sample can only be determined
with finite accuracy, leading to what often is called the Malm\-quist
edge-bias.  Assuming a symmetric error distribution on the selection
parameters (e.g.  diameter/magnitude), objects have an equal chance of
being scattered bins up as being scattered bins down.  Because there are
many more objects in the bins with smaller/fainter galaxies, on average
more objects are scattered up than down and we will overestimate the
number of objects in each bin.  We correct for this Malmquist selection
parameter uncertainty by calculating for each galaxy the weighted
average \vmax\ over the probability distribution of the selection
parameters.

%----------------------------------------------------------------------
\section{The nearby galaxy sample}
\label{sample}

We have used the sample described by Matthewson, Ford and Buchhorn
(\citeyear{MFB}, MFB hereafter) as the starting point for our
sample selection.  With more than a thousand field galaxies it is large
enough not to run immediately into low number statistics near the low
surface brightness and/or small scale size selection borders (de Jong \&
Lacey \citeyear{deJLac99}).  The main drawback of the sample is its
selection, as the sample was defined as a subsample from the ESO-Uppsala
Galaxy Catalog, which is a catalog selected by eye from photographic
plates. 

The MFB sample is in essence a diameter limited sample %
 %($1.65\arcmin\le D_{\rm maj}\le 5.05\arcmin$),
 of Sa-Sdm spiral galaxies and we can use Eq.\,(\ref{vmax}) for the
visibility correction.  Our reselection from the ESO-Uppsala Catalog
consists of 1003 galaxies, of which 860 having both MFB surface
photometry and redshifts.  The radial $I$-band luminosity profiles
provided by MFB were used to calculate total magnitudes (\magtot),
half total light (effective) radii (\re) and the average surface
brightnesses within these radii (\muave).  The 1D luminosity profiles
were decomposed in bulge and disk, using exponential light profiles for
both disk and bulge (method described in de Jong \citeyear{deJ2}), and
disk only structural parameters were derived. 

Distances were calculated from the redshifts using a Hubble constant of
65 Mpc\,km$^{-1}$\,s, corrected for peculiar velocities for those
galaxies included in the Mark III catalog (Willick et
al.~\citeyear{Wil97}).  The Galactic foreground extinction corrections
were calculated according to the precepts of Schlegel et
al.~(\citeyear{Sch98}).  We have used the internal extinction correction
method introduced by Byun (\citeyear{Byu92}, see also Giovanelli et
al.~\citeyear{Gio95}).  A parameter for which the
extinction correction has to be determined is first fitted against the
maximum rotation velocity of the disk ($V_{\rm rot}$) to reduce the
effect of distance dependent selection effects.  The residuals on this
fit are fitted against $\log(D_{\rm min}/D_{\rm maj})$ to empirically
determine the extinction as function of inclination.

%----------------------------------------------------------------------
\section{Space density distributions and a functional form}
\label{bivar}

\begin{figure}
\epsfxsize=0.85\textwidth
\epsfbox[0 144 526 700]{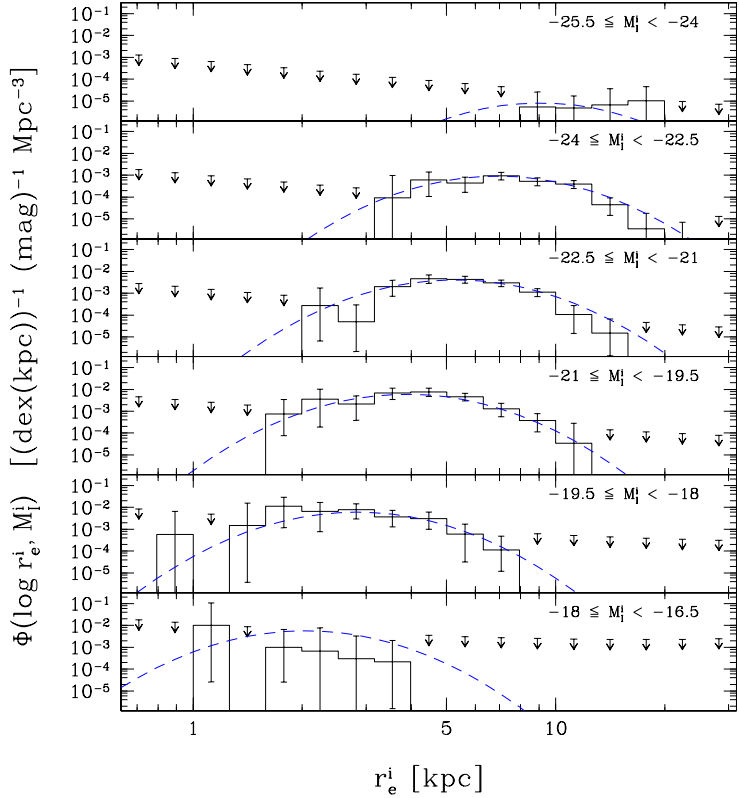}
\caption{The density distribution of Sa-Sm galaxies as
function of \magtot\ versus \re. The dashed line indicates the model
described by Eq.\,(\ref{bivareq})}
\label{bivar_magtot_re}
\end{figure}

Using the corrections described in
Sections\,\ref{viscor}\&\ref{sample}, we calculate the space
density of Sa-Sm galaxies in number of galaxies per Mpc$^3$ in the\linebreak
(\magtot,\re)-plane as shown in Fig.\,\ref{bivar_magtot_re}.  The
accuracy of the distribution is limited on the small scale length, low
luminosity end by selection effects (resulting in the large 95\%
confidence error bars), but these play hardly a r\^ole in determining
the distribution on the other end of the diagram.

There is great interest in deriving a functional form to describe the
observed bivariate distributions.  Parameterizations of the bivariate
distributions are useful when comparing distributions derived from
differently selected samples and when studying galaxy redshift
evolution.  The parameterization can also be used in modeling where both
galaxy luminosity and size are required (e.g.\ cross-section modeling of
Lyman forest host galaxies).  We will use here the bivariate
distribution in the (\magtot,\re)-plane, as this distribution is most
naturally connected to the galaxy formation scenario we are going to use
to find a functional form for the bivariate distribution.  Other parameter
combinations can easily be derived using $\magtot = \muave -
5\log(\re)-2.5\log(2\pi)$. 

To derive a parameterization of the bivariate distributions we will
follow the Fall \& Efstathiou (\citeyear{FalEfs80}) disk galaxy
formation theory.  Galaxies form in this theory in hierarchically
merging Dark Matter (DM) halos, giving rise to a distribution of DM halo
masses described by the Press \& Schechter (\citeyear{PreSch74}) theory,
which formed the inspiration for the Schechter LF (\citeyear{Sch76}). 
We will use a Schechter LF to describe the luminosity dimension of our
bivariate distribution function. 

In the Fall \& Efstathiou (\citeyear{FalEfs80}) model, the scale size of
a galaxy is determined by its angular momentum, which is acquired by
tidal torques from neighboring DM halos in the expanding universe.  The
total angular momentum of the system is usually expressed in terms of
the dimensionless spin parameter
$
\lambda = J |E|^{1/2} M_{\rm tot}^{-5/2} G^{-1}
$,
 with J the total angular momentum, $E$ the total energy and $M_{\rm
tot}$ the total mass of the system, all of which are dominated by the DM
halo.  N-body simulations (e.g.\ Warren et al.~\citeyear{War92}) show
that the distribution of $\lambda$ values acquired from tidal torques
can be approximated by a log-normal distribution with a dispersion
$\sigma_{\lambda}\sim 0.5-0.7$. 

A few simplifying approximations allow us to relate each of the factors
in the spin equation to our observed bivariate distribution parameters. 
A perfect exponential disk of effective size \re, (baryonic) mass $M_D$,
rotating with a flat rotation curve of velocity $V_{\rm rot}$ has
$J_D\propto M_D \re V_{\rm rot}$.  We assume that the specific angular
momentum of the disk is equal to the specific angular momentum of the
dark halo and therefore $J\propto M_D r_{\rm e} V_{\rm rot}$.  From the
virial theorem we get $E \propto V_{\rm rot}^2 M_{\rm tot}$.  We use a
power law between disk mass and luminosity: $M_D\propto L^\gamma$, with
$\gamma$ expected to be close to 1.  The $\gamma$ incorporates the
effect of $M_D/L$ variations due to gas mass fractions variations (de
Blok \& McGaugh~\citeyear{deBMcG97}) and stellar populations variations
(de Jong~\citeyear{deJ4};
Bell \& de Jong~\citeyear{BeldeJ99}) which tend to be a function of $L$. 
We take the
baryonic-to-DM ratio constant in each halo and assume that the same
fraction of the baryonic mass always ends up in the disk, resulting in
disk mass being proportional to total mass ($M_{\rm tot}\propto M_D$) in
each halo.  This is an assumption extremely hard to test, as nobody has
seen the `edge' of DM halos so far.  Now we only need the Tully \&
Fisher relation ($L\propto V_{\rm rot}^\epsilon$, with
$\epsilon\sim 3$ in the $I$-passband) to link the spin parameter
$\lambda$ to our observed bivariate distribution parameters. 
 
These approximations yield $\lambda \propto \re L^{(2/\epsilon-\gamma)}
\simeq \re L^{-1/3}$.  As $\lambda$ is expected to have a log-normal
behavior, this means that, {\em at a given luminosity, we expect the
distribution of scale sizes to be log-normal, and that the peak in the
\re\ distribution shifts with $\sim L^{-1/3}$}.  Combining this result
with the Schechter LF and using $\beta \equiv 2/\epsilon-\gamma$, the full
bivariate function for space density as function of luminosity and
effective radius becomes:
 \begin{eqnarray}
 \label{bivareq}
 \lefteqn{\Phi(\log\re,M)\,d\log\re\,dM =} \\
 &  & \frac{\Phi_0}{\sigma_\lambda \sqrt{2\pi}} 
     \exp(-\frac{1}{2}\left[\frac{\log\re/\res-0.4(M-M_*)\beta}
     {\sigma_\lambda/\ln(10)}\right]^2) \nonumber\\[1.3mm]
 &   & 0.4\ln(10) 10^{-0.4(M-M_*)(\alpha+1)} \,
     \exp(-10^{-0.4(M-M_*)}) \, d\log\re\,dM, \nonumber
 \end{eqnarray}
% \begin{eqnarray}
% \lefteqn{\Phi(\log\re,M)\,d\log\re\,dM =
%  \frac{\Phi_0}{\sigma_\lambda \sqrt{2\pi}} 
%     \exp(-\left[\frac{\log\re/\res-0.4(M-M_*)\beta}
%     {4\sigma_\lambda/\ln(10)}\right]^2)} \\[1.3mm] \nonumber
%&& 0.4\ln(10) 10^{-0.4(M-M_*)(\alpha+1)} \,
%     \exp(-10^{-0.4(M-M_*)}) \, d\log\re\,dM, %\nonumber
% \label{bivareq}
% \end{eqnarray}
 with the first line representing the log-normal scale size distribution
and the second line the Schechter LF in magnitudes ($M$).  In this
equation $\Phi_0$, $\alpha$ and $M_*$ have the usual meaning in a
Schechter LF. 

\begin{table*}
{\tabcolsep=1mm
\begin{tabular}{l|cccccc}
\hline
   & $\Phi_0$ & $\alpha$ & $M_*$ & \res & $\sigma_\lambda$ & $\beta$\\
%   & [mag$^{-1}$ dex$^{-1}$(kpc) Mpc$^{-3}]$ & & [$I$-mag] & [kpc] & & \\
\hline
total & 0.0031$\pm$0.0009 & -1.05$\pm$0.12 & -22.13$\pm$0.20 & 6.3$\pm$0.4 & 0.29$\pm$0.03 & -1/3$\pm$0\\
disk  & 0.0033$\pm$0.0008 & -1.04$\pm$0.11 & -22.30$\pm$0.18 & 6.1$\pm$0.4 & 0.37$\pm$0.03 & -1/3$\pm$0\\
\hline
total & 0.0033$\pm$0.0008 & -0.93$\pm$0.10 & -22.17$\pm$0.17 & 6.1$\pm$0.4 & 0.28$\pm$0.02 & -0.253$\pm$0.020\\
disk  & 0.0033$\pm$0.0007 & -0.90$\pm$0.10 & -22.38$\pm$0.16 & 5.9$\pm$0.3 & 0.36$\pm$0.03 & -0.214$\pm$0.025\\
\hline
\end{tabular}
 \caption{The results of fitting Eq.\,(\ref{bivareq}) to the
observed distributions.  Listed are the results for fits to the total
galaxy parameters and to the disk only parameters, either with $\beta$
fixed or free.  The errors are 95\% confidence limits.  $\phi_0$ is in
mag$^{-1}$ dex$^{-1}$(kpc) Mpc$^{-3}$, $M_*$ in $I$-mag, \res\ in kpc
and the other parameters are dimensionless}
 \label{fittab}
 }
\end{table*}

Before we can fit Eq.\,(\ref{bivareq}) to the data, we will have to
understand the uncertainty in the data points. The errors on the
\vmax\ corrected data points tend to be dominated by Poisson
statistics. Especially in bins where we have few galaxies, these
errors are highly asymmetric and we can not use a simple $\chi^2$
minimalization routine to fit the data. The 95\% confidence limits we
plot on the histogram of Fig.\,\ref{bivar_magtot_re} were calculated
taking both the distance uncertainty and the Poisson confidence limits
into account.  In a similar fashion we can account for bins with no
detections, calculating the Poisson probability distribution of a
non-detection for a exponential disk with given bin parameters and these
are indicated as upper limits in Fig.\,\ref{bivar_magtot_re}.

We used maximum likelihood fitting to determine the parameters in the
bivariate distribution function.  We used only the Poisson error
distribution to calculate the likelihood distribution on each bin, which
was minimized in the negative log.  We used bootstrap resampling to
estimate the errors on the bivariate distribution function parameters. 
Table\,\ref{fittab} lists the fit results for two cases, one for
\magtot\ and \re\ determined for the full galaxy (including bulge) and
one for the disk only.  Each case was fitted both with $\beta$ fixed to
$-1/3$ and $\beta$ free.  Surprisingly, the $\sigma_\lambda$ values of
$0.29-0.37$ are rather smaller than what is typically found for N-body
cosmological simulations. 

%Some possible explanations will be discussed in Section\,\ref{discuss}. 

%----------------------------------------------------------------------
\section{The Hubble Deep Field}

\begin{figure}
\epsfxsize=0.85\textwidth
\epsfbox[0 144 526 710]{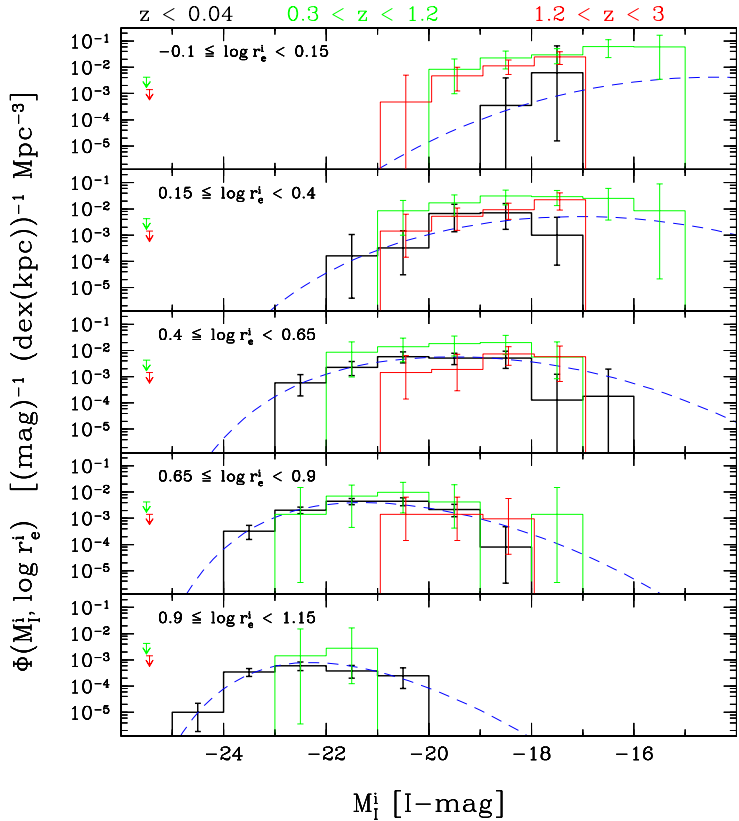}
 \caption{Redshift evolution of the bivariate density distribution of
 \magtot\ versus \re. The thick solid line $z<0.04$, light gray line
 $0.3<z<1.2$ and dark gray $1.2<z<3$. The dashed line indicates the
 model described by Eq.\,(\ref{bivareq})
 }
\label{hdf_magtot_re}
\end{figure}

With the Hubble space telescope we can now derive structural parameters
of high redshift galaxies.  Using the HDF (Williams~\citeyear{Wil96}) we
can estimate the bivariate space density of galaxies as function of
redshift using similar techniques as described before.  The situation is
slightly more complicated, because we have to take into account
cosmological corrections on the selection and derived parameters,
most notable the $(1+z)^4$ surface brightness dimming and the non-linear
decrease in diameter with redshift.  Furthermore we have to take
bandpass shift into account with K-corrections.  The derived \vmax\ of a
galaxy depends therefore on the assumed cosmology and galaxy evolution
model. 

We have used the structural galaxy parameters of Marleau \& Simard
(\citeyear{MarSim98}) and the photometric redshifts of Fern\'andez-Soto
et al.~(\citeyear{Fer99}) to derive to bivariate density distributions as
function of redshift from the HDF.  
 %We used non-evolving K-corrections
%and have taken \vmax\ corrections for both isophotal magnitudes and
%diameter limits into account. 
 Even though limited by low number statistics, the result in
Fig.\,\ref{hdf_magtot_re} suggests that there is only moderate density
evolution out to $z\sim 1.2$, but that the number density of bright,
large scale size galaxies decreases at higher redshifts.  The increase
in space density of the smallest galaxies ($\log\,\re<0.15$) at the
higher redshifts is probably due to the exclusion of Sm/Irr galaxies in
the local sample, which is not the case for the HDF sample.  This effect
is therefore probably not real or only true for late-type galaxies.

%----------------------------------------------------------------------
\section{Conclusions}
\label{discuss}

%Tully \& Verheijen (1997) have argued that the central surface brightness
%of galaxies shows a bimodal distribution, in particular when looking at
%$K$-band data.  We do not see such bimodality, independent whether we
%use their proposed bimodal dust extinction correction, we use only the
%200 most face-on galaxies with the smallest extinction correction, we
%use bulge/disk decomposed parameters or effective parameters.  In the
%many ways we have looked at the MFB data set, we have never seen any
%bimodality in the SB distributions.  Whether the bimodal effect is the
%result of the special Ursa Major cluster environment that was studied
%(even though a fair fraction of the MFB galaxies must lie in the outer
%parts of clusters) or an unlucky case of low number statistics (Bell \&
%de Blok ***) remains to be seen. 

% CDM models

We have quantified the local bivariate distribution function of spiral
galaxies.  It is well described by an Schechter LF in the luminosity
dimension and a log-normal distribution shifting by $L^{-\beta}$ in the
scale size dimension.  This parameterization gives an accurate
representation of the observed bivariate distributions, independently of
whether one believes in hierarchical galaxy formation models or in
CDM-like universes.  A detailed analysis of galaxy formation in CDM-like
universes paying attention to bivariate space density distributions will
appear in Cole et al.~(\citeyear{Col99}).

\bigskip
%\acknowledgements

\noindent
%We gratefully acknowledge Vince Ford, who provided the luminosity
%profiles for all MFB galaxies in electronically readable format.
Support for R.S.\ de Jong was provided by NASA through Hubble Fellowship
grant \#HF-01106.01-98A from the Space Telescope Science Institute,
which is operated by the Association of Universities for Research in
Astronomy, Inc., under NASA contract NAS5-26555.
%NED, ESO archive

\end{article}
\end{document}